\magnification=1200
\baselineskip=18truept

\input epsf

\def\preprint{Y}
\def\draftversion{N}

\if \draftversion Y


\fi

\def\figure#1#2#3{\vskip .2in
\if \preprint Y \midinsert \epsfxsize=#3truein
\centerline{\epsffile{figure_#1_eps}} \halign{##\hfill\quad
&\vtop{\parindent=0pt \hsize=5.5in \strut## \strut}\cr {\bf Figure
#1}&#2 \cr} \endinsert \fi}

\def\figureb#1#2{\if \preprint N \midinsert \epsfxsize=#2truein
\centerline{\epsffile{figure_#1_eps}} \halign{##\hfill\quad
&\vtop{\parindent=0pt \hsize=5.5in \strut## \strut}\cr \cr \cr
\cr \cr \cr  {\bf Figure #1} \cr} \endinsert \fi}


\def\npblong{1}
\def\recent{2}
\def\recentb{3}
\def\blumsoni{4}
\def\vranas{5}
\def\plbfirst{6}
\def\kaplan{7}
\def\shamir{8}
\def\seesaw{9}
\def\fn{10}
\def\taniguchi{11}
\def\taniguchib{12}
\def\yamada{13}
\def\npold{14}
\def\smitold{15}
\def\kikkaw{16}
\def\huetnn{17}

\line{\hfill KUNS-1485 HE(TH)97/21}
\line{\hfill RU-97-101}
\vskip 2truecm
\centerline{\bf Exponential suppression of radiatively induced mass in the
truncated overlap}
\vskip 1truecm

\centerline { Yoshio Kikukawa${}^a$, Herbert Neuberger${}^b$  
and Atsushi Yamada${}^c$\footnote{${}^{*}$}{On leave of absence from 
Department of Physics, University of Tokyo, Tokyo, 113, Japan.}.}
\vskip .5truecm

\centerline{${}^a$ Department of Physics}
\centerline{Kyoto University, Kyoto 606-01, Japan.}
\vskip .5truecm

\centerline {${}^b$ Department of Physics and Astronomy}
\centerline {Rutgers University, Piscataway, NJ 08855-0849, USA.}
\vskip .5truecm

\centerline {${}^c$ The Abdus Salam International Center 
for Theoretical Physics (ICTP)}
\centerline {Strada Costiera 11, P.O. Box 586, 34100, Trieste, Italy.}
\vskip 1.5truecm

\centerline{\bf Abstract}
\vskip 0.75truecm
A certain truncation of the overlap (domain wall fermions) 
contains $k$ flavors of Wilson-Dirac fermions. We show that
for sufficiently weak lattice gauge fields the effective mass
of the lightest Dirac particle is exponentially suppressed in $k$.
This suppression is seen to disappear when lattice topology is non-trivial.
We check explicitly that the suppression holds to one loop
in perturbation theory. We also provide a new expression for 
the free fermion propagator with an arbitrary additional mass term.
\vfill
\eject
{\centerline {\bf 1. Introduction}}
\vskip .5cm
The overlap exactly preserves 
the non-anomalous global chiral symmetries of vector-like
gauge theories [\npblong] on the lattice. Relatively to 
chiral gauge theories the overlap simplifies considerably
in the vector-like situation [\recent] 
but does not provide yet a practical alternative  
to more traditional simulations of QCD. The fundamental
reason for the simplification is that in the vector-like case
the Dirac operator can be viewed as a square matrix even
in topologically nontrivial backgrounds. 
For chiral theories the shape of the matrix can be 
square or rectangular depending on the gauge background
and therefore cannot be kept fixed: one needs an infinite
number of fermions. 
For vector like theories the rectangular shapes of the
two Weyl components always complement each other and can
be combined in one rigid square shape. As a result,
the overlap admits a truncation 
to a finite number of fermion fields ($k$). The
truncated overlap can be incorporated in practical 
simulations of QCD and holds the potential to become
competitive in the realm of small quark masses
[\recentb,\blumsoni,\vranas]. 
(The parameter $k$ can also be viewed as the length of an
extra dimension connecting two ``domain walls'' on 
which the Weyl components live [\plbfirst,\kaplan,\shamir].)
When $k$ is taken to infinity one regains the vector-like version of
the overlap where chiral symmetries are exact. At tree level,
the truncated overlap has an exponentially 
small (in $k$) quark mass. As stressed recently in [\recentb],
the mechanism for this suppression is quite generic. 
The generic features indicate that the suppression ought
to hold also after radiative corrections are taken into account.

But, the indication does not constitute proof even by normal
physics standards: It is 
worrisome that exact masslessness is protected
at infinite $k$ by an analytical index of a mass matrix, ${\cal M}$
[\plbfirst].
The index, $dim(ker({\cal M}^\dagger ))-dim(ker({\cal M} ))$ is present
at infinite $k$ (infinite number of flavors) and is robust under
small perturbations as any index would be. But, for any 
finite $k$, $dim(ker({\cal M}^\dagger ))= dim(ker({\cal M} ))$
and there is no index. 
Thus a non-smooth 
behavior as $k\rightarrow\infty$ cannot be easily excluded.
Chiral symmetries are notoriously difficult to maintain
on the lattice and we must be careful. 

If any of the good features of the overlap are to be even
partially preserved by the truncation we need new alternative
ways to understand
the suppression of the bare quark masses in the vector-like case,
without relying on continuity in $k$ at infinity. 
This prompted a search [\recentb] for other interpretations of
the mechanism: As was shown there, 
the mass suppression could be viewed as the outcome 
of a see-saw mechanism [\seesaw]
or, alternatively, of an approximate conservation law associated with a chiral
global symmetry in a way first devised by 
Froggatt and Nielsen [\fn]. In [\recentb] it was also 
found that a convenient link between finite and infinite $k$
is provided by a class of orthogonal polynomials 
associated with a certain measure on the real line. 
The effect of the
truncation is simply to keep only the first $k$ polynomials,
while the measure is $k$-independent. Therefore, the single
place $k$ enters is in the index of the last polynomial kept. 
The coefficients of the polynomials themselves are $k$-independent.
This makes it possible to trace the complete dependence on $k$.
Previous approaches [\shamir,\taniguchi,\taniguchib]
resorted to approximations in which
parts of the objects needed at finite $k$ were replaced by
their infinite $k$ limits. 

In the particular
model related directly to the overlap version employed
in [\recent] the polynomials
turned out to be the Chebyshev polynomials of second kind.
Other mass distributions would lead to other sets of orthogonal
polynomials but would work similarly. 

Let us summarize the situation: 
None of the arguments in favor of an exponential suppression at 
finite $k$ are fully compelling. The analytical index disappears,
and even continuity in $k$ is not a trivial matter. While the
orthogonal polynomials analysis does indicate continuity,
the exponential character of the 
suppression is evident only at tree level. 
The see-saw point of view more or less ignores doublers while
the F-N picture relies heavily on chiral symmetries, a notoriously
slippery concept on the lattice. Therefore, a direct check seems 
to be needed. 

As alluded above, we are not the first 
to undertake the task of checking
approximate masslessness in a direct way [\taniguchi,\taniguchib]. 
Actually, an even earlier calculation to one loop order for 
the exactly massless case can be found in [\yamada].
There, the perturbative calculation was carried out for the overlap
in the more general, chiral, case. 
This calculation essentially checked whether
the index argument survived one loop radiative corrections 
and concluded that it did. 

The work first announced in [\taniguchi] and more fully discussed 
in [\taniguchib] is still not complete, 
but it already claims to have  established that masslessness
is protected to one loop order.  
These authors ignore the relevance of the index at infinite $k$,
and this blurs one of the main effects of the truncation. 
It is troubling that the
calculation in [\taniguchib] does not use the exact free 
fermion propagators but
replaces them by propagators at $k=\infty$. The latter propagators
have one strictly massless quark and therefore extra infrared
singularities appear. Moreover, the  $k=\infty$ propagator ``knows''
about the analytic index at  $k=\infty$ and it is exactly 
the disappearance of the index at finite $k$ that makes the perturbative
test so relevant. In short, the replacement of the exact propagators
by the $k=\infty$ ones implies that certain $k$-dependent terms 
have been ignored and it is not proven that the ignored terms
are sub-dominant at large $k$.

A full computation is somewhat tedious.
We choose to avoid unnecessary details and focus on what the
systematics are. We wish to
know which features are essential for mass suppression
and which are not. At tree level, the mechanism of suppression
appears to be quite generic, and we would like to
know how much extra ``tweaking'' one could do to improve it even
further, without radiative corrections acting destructively. 
If the suppression indeed 
holds beyond tree level, it should do so for all kinds of
variants of the gauge action or the fermion-gauge
coupling. 
At tree level, even strict gauge invariance does not appear to be essential
for mass suppression. Therefore we expect the suppression to work
diagram by diagram (as long as we sum over all flavors in each -
so the diagram ``knows'' about the full flavor structure) and 
even before the momentum integration is done, i.e. directly at the
level of the Feynman integrand. 

Naively, the suppression appears quite miraculous.  
Consider first the basic
example of a single Wilson-Dirac
fermion on the lattice: In addition to the ``light'' Dirac fermion
there are ``heavy'' ones, the doublers. The gauge field can turn 
a light right handed Dirac fermion into a heavy right handed
doubler. The latter strongly mixes with its left handed heavy
partner which, again via the gauge interaction, can become 
a light left handed fermion. Thus, the light Dirac particle
gets a mass term of order $g^2$, where $g$ is the gauge coupling. 
There also is more direct source of mass in 
the gauge sea-gull coupling to the 
light fermion via the Wilson mass term.

Before considering several flavors let us 
ask why radiative mass generation is unavoidable. Probably
the most convincing form of the answer is the following:
If the right and left handed fermions did not mix we
would have exact $U(1)_V \times U(1)_A$ and we know this
cannot hold for all gauge fields because of instantons. 
Once we allow the helicities to mix, it appears impossible to assure
masslessness except by fine tuning (i.e. by adding a linearly
divergent mass counter-term - the divergence is ${\cal O} (a^{-1})$,
where $a$ is the lattice spacing). 

The truncated overlap contains a larger number ($16k -1$) 
of heavy Dirac fermions. Generically, this would not
invalidate the previous argument made for a single Wilson-Dirac
fermion and one would expect to have to fine tune. 
Strictly speaking, this is true for any finite $k$,
but in practice, any fine tuning (and the accompanying
chirality violating ${\cal O} (a)$ effects) 
could be ignored 
if the appropriate numerical coefficients 
were sufficiently small. 
Generically, one expects the coefficients to be of order one.
The exponential suppression of these coefficients 
is therefore potentially very useful,
but a bit mystifying. Some subtle cancelation must occur 
when one chooses the action in the way indicated by the overlap
or, equivalently, either by the see-saw or by the F-N mechanisms.

This paper has three main parts: In the first (next section) 
we show that in gauge backgrounds satisfying a
certain criterion 
there is exponential mass suppression. We also show that 
instantons evade this suppression by violating the criterion. 
Thus we have a separation of gauge backgrounds into two
classes, with approximate conservation of chiral symmetry
in one and order one breaking in the other. The first
class contains some neighborhood of the trivial
gauge orbit. This structure mirrors our understanding
in the continuum. However, 
this analysis does not shed light on 
the mysterious cancelation required in Feynman diagrams
containing internal gauge field lines. Nor is it
obvious to us that, even at one loop order, the gauge fields that
violate the criterion occur indeed with zero probability.  
In section 3 we complete the picture
by analyzing all the contributions to the zero momentum
fermion propagator, (in Feynman gauge)  to
one loop order in perturbation theory. The analysis reveals the 
cancelations between different flavors that are needed for the
suppression to hold, indicates that other sets of orthogonal
polynomials would also work just as they would at tree level
and shows, as expected, that highly nonlocal
pure gauge interactions could spoil the suppression. 

Mainly for future use we present in 
section 4 the structure of the free propagator 
in a way that should generalize to other 
sets of orthogonal polynomials. The full expression for the
Chebyshev case (truncated overlap) is also given. 
Some of the technical details are relegated to an appendix.
Our derivation is different from the one adopted in [\plbfirst]
(and later generalized to the truncated case in [\shamir,\taniguchib])
and leads to a particular form of presenting the result that
is both more concise and, we feel, more insightful. 
We make no direct use of these results here, but
they should be useful in other situations. 

The final section contains a summary and outlook.
\vskip 1cm

\centerline{\bf 2. Light quark propagator in various gauge backgrounds.}
\vskip .5cm
First, we establish our  notation. It is very similar to that of [\recentb].
The action $S$ (appearing with a minus sign in the exponent when
used to define the partition function) is:
$$
S={1\over{2g^2}} S_g +S_F +S_{pf}.\eqno{(2.1)}$$
$S_g$ is a pure gauge action whose detailed form is irrelevant here.
The other two terms contain the fermions as Grassmann
variables ($S_F$) and the pseudo-fermions as normal numbers,
($S_{pf}$). We use both for the fermions and pseudo-fermions the
following left-right structure:
$$
\pmatrix{ \Phi_1 \cr \Phi_2 \cr \vdots\cr\vdots\cr\Phi_{2k-1}\cr
\Phi_{2k}\cr} =
\pmatrix{ \chi_1^R \cr \chi_1^L \cr \vdots\cr\vdots\cr\chi_{k}^R\cr
\chi_{k}^L\cr} .\eqno{(2.2)}$$
$\chi_j^{R,L}$ are left or right Weyl fermions in the notation of
[\npold].
Similarly one defines $\bar\Phi_s$. Our convention is
that vector gauge interaction appear diagonal
in $D$. We suppressed all space-time, spinorial and 
gauge indices, displaying
explicitly only the left-right character and flavor.

The lattice is taken to have $L^4$ sites. Our basic building blocks in the
lattice Dirac matrix $D$ will have size $q\times q$ where, 
in 4 dimensions, $q=2 n_c L^4$. $n_c$ is the dimension of 
the gauge group representation ($n_c =3 $
for QCD). Following [\npold] and [\recentb] we write:
$$
D=\pmatrix{ C^\dagger & B & 0 & 0 & 0 & 0 &\ldots &\ldots & 0 & 0\cr
            B 	     & -C & -1& 0 & 0 & 0 &\ldots &\ldots & 0 & 0\cr
	    0   &-1& C^\dagger& B & 0 & 0 &\ldots &\ldots & 0 & 0\cr
            0 & 0  &     B    &-C &-1 & 0 &\ldots &\ldots & 0 & 0\cr
	    0 & 0  & 0 &-1& C^\dagger & B &\ldots &\ldots & 0 & 0\cr
            0 & 0  & 0 & 0&  B        &-C &\vdots &\ldots & 0 & 0\cr
            \vdots&\vdots&\vdots&\vdots&\vdots&\ddots&\ddots&\ddots
&\vdots&\vdots\cr
            \vdots&\vdots&\vdots&\vdots&\vdots&\vdots&\ddots&\ddots
&\ddots&\vdots\cr
	    \ldots&\ldots&\ldots&\ldots&\ldots&\ldots&\ldots&\ddots
&\ddots&\ldots\cr
	    0 & 0 &  0 & 0& 0         & 0 &\ldots &\ldots & B &-C\cr}.
\eqno{(2.3)}$$
The matrix $D$ is of size $2k\times 2k$ and the entries 
are $q\times q$ blocks. 

The matrices $B$ and $C$ are dependent on the gauge background defined
by the collection of link matrices $U_\mu (x)$. The link matrices
are of dimension $n_c\times n_c$. $\mu$ labels the $4$ positive directions
on a hypercubic lattice and  $U_\mu (x)$ is the unitary matrix associated
with a link that points from the site $x$ in the $\hat \mu$-direction. 
$$
\eqalign{
( C )_{x \alpha a, y \beta b}
& ={1\over 2} \sum_{\mu=1}^{4} \sigma_\mu^{\alpha\beta}
[\delta_{y,x+\hat\mu} (U_\mu (x) )_{ab} -
\delta_{x,y+\hat\mu} (U_\mu^\dagger (y))_{ab}] \equiv 
\sum_{\mu=1}^{4} \sigma_\mu^{\alpha\beta} ( W_\mu )_{xa,yb}~ .\cr
( B_0 )_{x \alpha a, y \beta b} & = {1\over 2} \delta_{\alpha\beta} 
\sum_{\mu =1}^{4} [2\delta_{xy}
\delta_{ab} - \delta_{y,x+\hat\mu} (U_\mu (x) )_{ab} -
\delta_{x,y+\hat\mu} (U_\mu^\dagger (y))_{ab} ].\cr
( B )_{x \alpha a, y \beta b} & =( B_0 )_{x \alpha a, y \beta b} 
+ 
M_0 \delta_{x\alpha a,y\beta b}.\cr}\eqno{(2.4)}$$
The indices $\alpha,\beta$ label spinor indices and take values $1$ 
or $2$. The indices $a,b$ label color in 
the range $1$ to $n_c$. The Euclidean $4 \times 4$
Dirac matrices $\gamma_\mu$
are taken in the Weyl basis where their form is
$$
\gamma_\mu = \pmatrix {0&\sigma_\mu \cr \sigma_\mu^\dagger & 0\cr}.
\eqno{(2.5)}$$

As long as the parameter $M_0$ satisfies $0<M_0 < 1$, $B$ is positive
for any gauge field background and the light tree level quark mass $m_{RL}$
is small [\recentb,\vranas]:
$$
m^2_{RL}= |M_0|^{2k} (1-|M_0|^2 )^2 (1+{\cal O}( |M_0|^{2k} )) .
\eqno{(2.6)}$$

We have
$$
S_F =-\sum_{s=1}^{2k} \bar\Phi_s (D\Phi )_s .\eqno{(2.7)}$$
and 
$$
S_{pf} = -\sum_{s=1}^{2k} \bar\Phi_s^{pf} 
(D^{pf}\Phi^{pf} )_s .\eqno{(2.7)} $$
where
$$
D^{pf}=\pmatrix{ C^\dagger & B & 0 & 0 & 0 & 0 &\ldots &\ldots & 0 & 1\cr
            B 	     & -C & -1& 0 & 0 & 0 &\ldots &\ldots & 0 & 0\cr
	    0   &-1& C^\dagger& B & 0 & 0 &\ldots &\ldots & 0 & 0\cr
            0 & 0  &     B    &-C &-1 & 0 &\ldots &\ldots & 0 & 0\cr
	    0 & 0  & 0 &-1& C^\dagger & B &\ldots &\ldots & 0 & 0\cr
            0 & 0  & 0 & 0&  B        &-C &\vdots &\ldots & 0 & 0\cr
            \vdots&\vdots&\vdots&\vdots&\vdots&\ddots&\ddots&\ddots
&\vdots&\vdots\cr
            \vdots&\vdots&\vdots&\vdots&\vdots&\vdots&\ddots&\ddots
&\ddots&\vdots\cr
	    \ldots&\ldots&\ldots&\ldots&\ldots&\ldots&\ldots&\ddots
&\ddots&\ldots\cr
	    1& 0 &  0 & 0& 0         & 0 &\ldots &\ldots & B &-C\cr}.
\eqno{(2.9)}$$

A convenient choice for interpolating fields for the light quark 
([\plbfirst]) are the 
quantities $\chi_1^R , \chi_k^L ~(\Phi_1 , \Phi_2 )$
To obtain their correlation functions we introduce two 
source terms, $X$ and $Y$, in $D$ [\recentb]. 
$$
D(X,Y)=\pmatrix{ C^\dagger & B & 0 & 0 & 0 & 0 &\ldots &\ldots & 0 & X\cr
            B 	     & -C & -1& 0 & 0 & 0 &\ldots &\ldots & 0 & 0\cr
	    0   &-1& C^\dagger& B & 0 & 0 &\ldots &\ldots & 0 & 0\cr
            0 & 0  &     B    &-C &-1 & 0 &\ldots &\ldots & 0 & 0\cr
	    0 & 0  & 0 &-1& C^\dagger & B &\ldots &\ldots & 0 & 0\cr
            0 & 0  & 0 & 0&  B        &-C &\vdots &\ldots & 0 & 0\cr
            \vdots&\vdots&\vdots&\vdots&\vdots&\ddots&\ddots&\ddots
&\vdots&\vdots\cr
            \vdots&\vdots&\vdots&\vdots&\vdots&\vdots&\ddots&\ddots
&\ddots&\vdots\cr
	    \ldots&\ldots&\ldots&\ldots&\ldots&\ldots&\ldots&\ddots
&\ddots&\ldots\cr
	    Y & 0 &  0 & 0& 0         & 0 &\ldots &\ldots & B &-C\cr}.
\eqno{(2.10)}$$
$X$ and $Y$ are $q\times q$ matrices which act as arbitrary sources for the
bilinears $\bar\chi_{1\alpha a}^R (x) \chi_{k\beta b}^L (y) $,
$\bar\chi_{k\alpha a}^L (x) \chi_{1\beta b}^R (y) $. To obtain the correlation
functions we need to expand the ratio ${{\det D(X,Y)}\over {\det D^{pf}}}$
around $X=0$, $Y=0$ to linear order in $X$ and $Y$.
Actually, $D^{pf} = D(X={\bf 1}, Y={\bf 1})$.  In the appendix of
[\recentb] the following identity was established:
$$
{{\det D(X,Y)}\over {\det D^{pf}}}= 
{{\det 
\left [ \pmatrix {-X &0 \cr 0& 1 } - T^{-k} 
\pmatrix {1&0\cr 0 &-Y\cr } \right ] }\over
{\det \left [ 1+T^{-k} \right ]} }.\eqno{(2.11)}$$
Here:
$$
T=\pmatrix {{1\over B} & {1\over B} C \cr
         C^\dagger {1\over B} & C^\dagger {1\over B} C +B \cr}.
	 \eqno{(2.12)}$$
$T$ is positive definite for any gauge field and $\det T \equiv 1$.
Defining the hermitian matrix $H$ by
$$
T=e^{-H}.\eqno{(2.13)}$$
we see that $tr (H)=0$. 

The spectra of $H$ and $T$ are gauge invariant. For a background $U_\mu (x)$
gauge equivalent to  $U_\mu (x)\equiv 1$, $H$ has $q$ positive
eigenvalues and $q$ negative ones. Around zero, $H$ has a finite gap.
Let us write $U_\mu (x)=e^{igA_\mu (x)}$ where the $A_\mu (x)$ are fixed
by some local gauge condition (Feynman gauge - we have no
problems with Gribov copies here as we simply assume that $A_\mu (x)$
satisfies the gauge condition to identify the orbit). We see that
the structure of the spectrum could be preserved in some neighborhood
of $g=0$ if the probability distribution for $ A_\mu (x)$ 
(at leading order this probability distribution is $g$ independent)
suppresses large $A_\mu (x)$ strongly enough. Link
configurations causing $H$ to have 
equal numbers of positive and negative
eigenstates, and with the minimal absolute value 
eigenvalue $e_{\rm min}$ satisfying $e_{\rm min} k >> 1$, 
are referred to as satisfying
the ``perturbative'' criterion for $k$. For any $k$ one can find
a small enough range in $g$ that a typical link will satisfy
the perturbative criterion with high probability but it is
difficult to make this statement more precise. This difficulty
is the main reason for also performing a direct one loop calculation
in the next section. 

To compute the two determinants in eq. (2.11) we use two different 
orthonormal bases, and represent the matrices
by their matrix elements between the two bases. 
This trick produces expressions that are
equivalent in form to those of the overlap ([\npblong]), but
now with the truncation effects made explicit.
The first basis is denoted generically by $v$ and is gauge field
independent. $v$ is indexed by a pair of indices: the first is
either $r$ or $l$ while the second is $J=1,2,....q$. Each $J$
is a short-hand for a triplet of indices: a Weyl index, a lattice site
and a gauge group index ($J=(\alpha , x, a)$). 
$$
v^{(r,J)} = \pmatrix { v^J \cr 0\cr},~~~~~~~~
v^{(l,J)} = \pmatrix { 0 \cr v^J \cr},~~~~~v^J_{\beta y b}
=\delta_{\alpha\beta}\delta_{ab}\delta_{xy}.\eqno{(2.14)}$$
The other basis, generically denoted by $w$, is made out of 
ortho-normalized eigenvectors of $H$. 
These vectors do depend on the gauge fields
and transform covariantly under gauge transformations. 
Without restricting generality we can take the gauge group as $SU(N_c )$,
and therefore, once the bases are used for evaluating determinants,
the gauge dependence disappears on account of the unimodularity
of the group elements and the ultra-local structure of the basis $v$. 

We assume that the background satisfies our perturbative criterion.
The $q$ positive eigenvectors of $H$ are denoted 
by $w^{(p,J)}$, $J=1,2,...q$ and the $q$ negative ones by $w^{(n,J)}$:
$$\eqalign{
Hw^{(p,J)}& = E^p_J w^{(p,J)}\cr
Hw^{(n,J)}& = -E^n_J w^{(n,J)}.\cr}\eqno{(2.15)}$$

The $2q\times 2q$ 
unitary matrix relating the $v$ and the $w$ bases is given by:
$$
{\bf U} =\pmatrix {w^{(p,I)\dagger} v^{(r,J)}&
w^{(p,I)\dagger} v^{(l,J)}\cr
w^{(n,I)\dagger} v^{(r,J)}&
w^{(n,I)\dagger} v^{(l,J)}\cr } 
=\pmatrix {U^{pr} & U^{pl}\cr U^{nr} & U^{nl}\cr}.
\eqno{(2.16)}$$
We choose the phases so that $\det U^{pr} = (\det U^{nl})^*$
(this is possible since the unitarity of $\bf U$ 
implies $|\det U^{pr}| = |\det U^{nl}|$, as shown in [\npblong]).
With this choice $\det{\bf U} =1$. For any $2q \times 2q$
matrix $Z$ it is true then 
that $\det Z = \det_{(\xi I ), (\eta J )} w^{(\xi I)\dagger} Z v^{(\eta J)}$ 
where $\xi = p,n$ and $\eta=l,r$.

Applying this to our basic expressions, we obtain:
$$
\eqalign{
&\det 
\left [ \pmatrix {- X &0 \cr 0& 1 } - T^{-k} 
\pmatrix {1&0\cr 0 &- Y\cr } \right ] =\cr
&
\det\pmatrix { -w^{(p,I)\dagger} \hat X v^{(r,J)}-e^{kE^p_I} 
w^{(p,I)\dagger} v^{(r,J)} & 
e^{kE^p_I} w^{(p,I)\dagger} \hat Y v^{(l,J)} + w^{(p,I)\dagger} v^{(l,J)}\cr
-w^{(n,I)\dagger}\hat X v^{(r,J)}-e^{-kE^n_I} w^{(n,I)\dagger} v^{(r,J)} &
e^{-kE^n_I} w^{(n,I)\dagger}\hat Y v^{(l,J)}+
w^{(n,I)\dagger} v^{(l,J)} 
\cr}.\cr}\eqno{(2.17)}$$
Here, $\hat X = \pmatrix { X & 0\cr 0 & X\cr}$ and $\hat Y = \pmatrix
{Y & 0\cr 0& Y }$.

As $k\rightarrow \infty$ we can write:
$$\eqalign{
&
\det \left [ \pmatrix {-X &0 \cr 0& 1 } - T^{-k} 
\pmatrix {1&0\cr 0 &-Y\cr } \right ] \sim\cr
&
\det\Biggl\lbrace
\pmatrix { e^{kE^p_I} \delta_{IJ} & 0\cr 0 & \delta_{IJ}\cr}
\left [ \pmatrix {-w^{(p,I)\dagger} v^{(r,J)}&
w^{(p,I)\dagger} {\hat Y} v^{(l,J)}\cr
-w^{(n,I)\dagger}{\hat X} v^{(r,J)}&
w^{(n,I)\dagger} v^{(l,J)}\cr } +{\cal O} (e^{-ke_{\rm min}})\right ]
\Biggr\rbrace .\cr}\eqno{(2.18)}$$
On the other hand, we have, for large $k$,
$$
\det [ 1+T^{-k} ] \sim \det 
\pmatrix { e^{kE^p_I} \delta_{IJ} & 0\cr 0 & \delta_{IJ}\cr}.\eqno{(2.19)}$$
where we now used only the $w$ basis on both sides. At large $k$ we
end up with
$$
{{\det D(X,Y)}\over {\det D^{pf}}}\sim
\det \pmatrix {-w^{(p,I)\dagger} v^{(r,J)}&
w^{(p,I)\dagger} \hat Y v^{(l,J)}\cr
-w^{(n,I)\dagger}\hat X v^{(r,J)}&
w^{(n,I)\dagger} v^{(l,J)}\cr }.\eqno{(2.20)}$$

Setting $X=Y=0$ we obtain the overlap result [\npblong]:
$$
{{\det D(X=0,Y=0)}\over {\det D (X={\bf 1},Y={\bf 1})}}=
|\det U^{pr}|^2 = |\det U^{nl}|^2 .\eqno{(2.21)}$$

Just like in [\recentb], our derivation made no use of
fermion creation and annihilation operators. The latter
are essential for chiral gauge theories, and,
as a consequence, can be also used in 
the vector-like case [\npblong]. But operators techniques
become quite clumsy when dealing with the truncation of the
overlap, and unnecessarily complicate the derivations. 

Now, we need to expand to linear order in $X$ and $Y$. But,
if we replace any of $X$ or $Y$ by zero, the dependence on the
other variable drops out. We therefore conclude that, up to
corrections exponentially small in $k$, 
$<\chi^R_{1\alpha a} (x) \chi^L_{k\beta b}(y)>_U$ and
$<\chi^L_{k\alpha a} (x) \chi^R_{1\beta b}(y)>_U$ vanish.
The same derivation would show that, 
if the number of positive and negative eigenvalues differs
by two say, the above correlation functions do not have to vanish.
Note that it is only one part of the perturbative criterion that
gets violated. As long as the second part of the criterion is
satisfied, namely an appropriately gapped spectrum, the large
$k$ limit, be it zero or nonzero, is approached exponentially
fast. One could say that a weaker 
criterion, namely one which contains only the gap
requirement, defines the subset of configurations to
which semi-classical considerations apply [\recentb]. 

As already mentioned, it is difficult to make precise probabilistic
estimates on how often we have at most $e_{\rm min} k \sim 1$ 
for any given $k$
and $g$. Therefore, the result of a perturbative calculation of
the light quark propagator in Feynman gauge, even
at one loop, is not predicted with certainty by the result of the
present section.
Had we been able to claim that gauge fields (assumed {\it a priori}
to produce $H$'s having equal numbers 
of positive and negative eigenvalues),
violating the $e_{\rm min} k >>1$ for a given $k$, occur with
vanishing probability as $g$ is smaller than some 
small, $k$-independent,  $\epsilon >0$, we could conclude, just
on the basis of this section, that exponential mass suppression
holds to all orders in perturbation theory. 
In the absence of this claim,
the main importance of the present result is in establishing an
effective decoupling between the left and right components of
the light quark in ``perturbative'' backgrounds while 
also showing specifically how this decoupling gets spoiled
in other backgrounds. 
\vskip 1cm
\centerline{\bf 3. Exponential mass suppression to one loop.}
\vskip .5cm

The infinite mass matrix ${\cal M}$ was introduced
in [\plbfirst]. As emphasized there, although
one wrote down a theory that
{\sl formally} looked vector-like, due to the impossibility
of rotating all the fields so as to make ${\cal M}$ hermitian, the theory  
ended up being chiral. Of course, this slightly  paradoxical 
situation occurred only because of the infinite dimensionality
of ${\cal M}$. With infinite mass matrices inherently
non-hermitian, the free fermion propagator
had to be separated into left and right parts, and separate expressions
had to be written for each. This approach has been taken over to
the vector-like case in [\taniguchi] and [\taniguchib]. 
Clearly, in the vector-like case it constitutes an unnecessary
complication, particularly in the truncated overlap situation, 
where it should be transparently clear that all we 
are dealing with are several lattice Dirac fermions mixed in a special
way. 

So, following [\recentb], we change bases in flavor space to make the
mass matrices hermitian. At the beginning, to avoid confusion,
we shall denote explicitly the direct product structure between Dirac
spinor space and the rest. Writing out the explicit index dependence
on flavors ($i,j=1,2...k$) we have, in the new flavor basis [\recentb], 
$$
D_{ij} = \gamma_\mu \otimes W_\mu ~ \delta_{ij} + 1 
\otimes M_{ij},\eqno{(3.1)}
$$
where the $W_\mu$ were defined in (2.4) and a sum over $\mu$ is implied.
The hermitian matrix $M$ has the following flavor structure:
$$
M=\pmatrix{0&0&\ldots&0&-1&B\cr
	   0&0&\ldots&-1&B&0\cr
	   \vdots&\vdots&\vdots&\vdots&\vdots&\vdots\cr
	   \vdots&\vdots&\vdots&\vdots&\vdots&\vdots\cr
	   -1&B&\ldots&\ldots&0&0\cr
	   B&0&\ldots&\ldots&0&0\cr}.\eqno{(3.2)}$$
The entries are $n_c L^4 \times n_c L^4$ matrices. 

To set up perturbation theory, set $U_\mu (x)\equiv 1$ 
and go to Fourier space. The site dependence of $M$
can be diagonalized and each entry becomes block
diagonal with momentum ($p$) dependent $n_c\times n_c$
blocks. There are $L^4$ such blocks in each entry.
Actually, the factor in group space is unity, so we
simply have each block of $M$ represented by one real
function of momentum. Let the matrices made out of these
representatives, now of dimension $k\times k$, be denoted
by $m(p)$. 

For any $p$
the matrix $m^2 (p)$ is tridiagonal and has been diagonalized
in [\recentb]. Let us summarize what we need
here. The eigenvalues are denoted by $\mu^2_s (p)$
and are given by
$$
\mu^2_s (p) = 1+b^2 (p) -2b(p)\lambda_s (p),~~~~~~~b(p)\equiv \sum_\mu
(1-\cos p_\mu ) +M_0 , \eqno{(3.3)}$$
where the $\lambda_s (p)$ are the k real roots of the polynomial
equation
$$
U_k (\lambda )=b(p) U_{k+1} (\lambda ).\eqno{(3.4)}$$
The $U_j (\lambda )$ are the
Chebyshev orthogonal polynomials of the second kind and play a central
role in what follows. We are quite convinced that other mass matrices,
associated 
with other sets of orthogonal polynomials, would work similarly and in
order to stress that we shall try to avoid using any specific
properties of the Chebyshev polynomials. The dependence on 
the momentum $p$ only enters through $b(p)$, but this could easily
be changed in other variants. 

The eigenvectors corresponding
to the above eigenvalues are proportional to $U_j (\lambda_s (p))$:
$$
\sum_{j=1}^k m^2 (p)_{ij} U_j (\lambda_s (p)) =
\mu^2_s (p) U_i (\lambda_s (p)).\eqno{(3.5)}$$
The orthonormal eigenvectors are given by
$$
O_j (\lambda_s (p) )= N (\lambda_s (p) )  U_j (\lambda_s (p))
\equiv N_s (p) U_j (\lambda_s (p)),\eqno{(3.6)}$$
where, for any $\lambda$,
$$
N^2 (\lambda ) = {1\over {\sum_{j=1}^k U^2_j (\lambda )}}.
\eqno{(3.7)}$$
For definiteness, we choose to define $N(\lambda)$ as the positive
square root of (3.7). 

The orthogonal character of the polynomials ensures that
all the $\lambda_s (p)$ are distinct for any fixed $p$.
We can arrange the $\lambda_s (p)$ in descending order,
so that $s=1$ corresponds to the smallest mass. 
For each $s$, $\lambda_s (p)$ is a smooth function of $p$,
as no crossings can occur. The associated set 
of $k$ eigenvectors are globally and smoothly 
defined over momentum space.
The lack of degeneracy for any $p$
implies that diagonalizing $m^2 (p)$ has
also diagonalized $m (p)$. Therefore,
$$
\sum_{j=1}^k m_{ij} (p) O_j (\lambda_s (p)) = 
\mu_s (p) O_i (\lambda_s (p) ) .\eqno{(3.8)}$$
The above equation {\sl defines} the sign of $\mu_s (p)$. More explicitly
it can be read off from
$$
\mu_s (p) ={{b(p)}\over {U_k (\lambda_s (p))}}=
{1\over {U_{k+1} (\lambda_s (p))}},\eqno{(3.9)}$$
or from
$$
\mu_s (p) = \left [ {1\over{b(p)}} + b(p) -2\lambda_s (p) 
\right ] U_k (\lambda_s (p)).\eqno{(3.10)}$$
$\mu_s (p)$ has the same sign as $U_k (\lambda_s (p))$. 
From the above two relations one also derives
$$
1+b^2 (p) -2b(p)\lambda_s (p) = \left [ {{b(p)}\over {U_k (\lambda_s (p))}}
\right ]^2 ,\eqno{(3.11)}$$
in agreement with (3.3). 

It was shown in [\recentb] that, as long as
$$
0\le b(p)<1+{1\over k},\eqno{(3.12)}$$
$\lambda_1 (p)$ is larger than unity.  
Let the region ([\plbfirst]) of momentum
space where $0\le b(p)\le 1+{1\over k}$ 
be denoted by $R$. For $p\in R$ we have,
as $k\to\infty$ ([\recentb]), 
and up to exponentially suppressed corrections,
$$
\lambda_1 (p) \sim {1\over 2} \left [ b(p) + {1\over{b(p)}}\right ],
~~~~~~~~\mu_1^2 (p) \sim b^{2k}(p) [1-b^2(p) ]^2 .\eqno{(3.13)}$$
For all $2\le s \le k$ with any $p$ and also for $s=1$ 
with $p \not\in R$ $|\lambda_s (p)|\le 1$ 
and $\mu_s (p)$ stays finite and nonzero. Thus, the masses of all
the heavy (doublers and extra flavors)
$16k-1$ Dirac fermions are given by the ultra-violet
cutoff times order one coefficients. The corresponding eigenvector
components, $O_j (\lambda_s (p))$, are of typical 
order ${1\over\sqrt{k}}$
and have, in general, an oscillatory behavior as a function of $j$. 
The normalization constants $N_s (p)$ are also of 
order ${1\over\sqrt{k}}$ in these cases.

However, for $s=1$ and $p\in R$ , where the light quark state resides,
the components $O_j (\lambda_1 (p))$ do not oscillate but rather
vary exponentially for large $j$ (we assume $k$ is large). 
Similarly, $N_1 (p)$ is exponentially large in that region. All this
is a direct consequence of the exponential growth of the
polynomials $U_j (\lambda )$ outside their interval
of orthogonality ($|\lambda|\le 1$) and their boundedness within.
This property will be shared by other sequences of orthogonal
polynomials, at least as long as the interval of orthogonality
is a finite segment. 

The above information is basically
all we wish to use when analyzing the light 
fermion propagator to one loop.
In momentum space  the full free propagator 
is given by:
$$
G^{(0)} (p) = {{-i\gamma_\mu \otimes \bar p_\mu {\bf 1} + 1\otimes m(p)}
\over {1\otimes (\bar p^2 {\bf 1} + m^2 (p))}},
~~~~~~~~\bar p_\mu = \sin (p_\mu ).\eqno{(3.14)}$$
Note that the matrices in the numerator and the denominator
commute so there is no ordering problem and (3.14) is unambiguous. 
We ignored the trivial color dependence. Introducing the
diagonal $k\times k$ matrices $\mu(p) $, with $\mu_1 (p),
\mu_2 (p), ...\mu_k (p)$ along the diagonal and the orthogonal
matrices $O(p)$ made out of the eigenvectors
$O_j (\lambda_s (p))$ as columns, we can write the
free propagator as
$$
G^{(0)}(p) = O(p) {{-i\gamma_\mu \bar p_\mu +  \mu(p)}
\over { \bar p^2  + \mu^2 (p)}} O^T (p).\eqno{(3.15)}$$
Above, we dropped the explicit direct products and anything that is unity
in the appropriate space. 

We now proceed with the calculation of the fermion propagator to one loop.
As mentioned in the introduction we only wish to 
set things up and establish exponential suppression
of the numerical coefficient of
the ${\cal O }(a^{-1})$ radiatively induced mass for $s=1$
and $p\in R$, where $a$ is
the lattice spacing. We shall avoid any explicit calculation that
is tangential to our goal. We choose to define the quark mass from
the expansion of the inverse propagator around zero four momentum.
Since all fermions are massive this expansion is not infrared
divergent.

Expanding (3.1) to order $g^2$ we have:
$$
D_{ij}=D^{(0)}_{ij} + M^{(0)}_{ij} + g \gamma_\mu W_\mu^{(1)} \delta_{ij}
+g M^{(1)} P_{ij} + g^2 \gamma_\mu W_\mu^{(2)} \delta_{ij} + g^2
M^{(2)} P_{ij}.\eqno{(3.16)}$$
Here, $P_{ij} = \delta_{i,k+1-j}$ implements physical parity exchanging
the left and right components of the light fermion. 

The propagator, $G\equiv {1\over D}$, is also expanded to order $g^2$,
and after that averaged over the gauge fields with a Gaussian measure, 
using Feynman gauge. The order $g$ term averages to zero and we can write:
$$
G={1\over{D^{(0)} - g^2  {\bf \Sigma}}},\eqno{(3.17)}$$
where the self energy ${\bf \Sigma}$ is given by
$$
{\bf \Sigma}=\langle D^{(1)} {1 \over { D^{(0)} } } 
D^{(1)} - D^{(2)} \rangle ,
\eqno{(3.18)}$$
with $\langle ... \rangle$ denoting gauge averaging. In (3.18)
we used the following short-hand notations:
$$\eqalign{
D^{(1)}_{ij} = &\gamma_\mu W_\mu^{(1)} \delta_{ij} + M^{(1)} P_{ij}\cr
D^{(2)}_{ij} = &\gamma_\mu W_\mu^{(2)} \delta_{ij} + M^{(2)} P_{ij}.\cr}
\eqno{(3.19)}$$

We are interested in the self energy in momentum space, 
near zero momentum. Since we wish to find out the mass shift
of the lightest mode (labeled by $s=1$), we only need the diagonal 
matrix element of the self energy
in the unperturbed light flavor eigenstate, whose
wave function in flavor space is given by $O_j (\lambda_1 (p))$. 

First we wish to establish that we are not really
interested in the {\it vicinity} of zero momentum, since all we would get
from it is the wave function renormalization constant ${\cal Z}$. 
By showing that ${\cal Z}$ cannot
be exponentially large in $k$, 
so it cannot affect our
conclusion about the possible exponential suppression of the light
quark radiatively-induced mass, we can restrict 
our analysis to the self energy strictly at zero momentum.
That ${\cal Z}$ cannot be exponentially large in $k$ is quite obvious:
The matrix $O$ in (3.15) is orthogonal so its entries are bounded.
There are no divergences worse than logarithmic in the infinite $k$
limit (when one of the quarks becomes massless). 
Thus, at most, ${\cal Z}$ could have a linear dependence on $k$, and
we can forget about ${\cal Z}$ altogether.

The $\langle D^{(2)}\rangle$ (``tadpole'') 
term has no free propagators. The factor 
containing $W_\mu^{(2)}$ has no left-right terms (terms
commuting with $\gamma_5$) so does not
contribute. To estimate the contribution of $M^{(2)}$ we only need
to compute the expectation value of $P$ in the $s=1$ state. The structure
of $M^{(2)}$ is irrelevant, since all it determines is a multiplicative
factor of order one. 

The following identity follows directly from the recursion relations
for the Chebyshev polynomials:
$$
\sum_{i=1}^k U_i (\lambda ) U_{k+1-i} (\rho )=
{1\over 2} {{U_{k+1} (\lambda ) - U_{k+1 } (\rho )}
\over {\lambda-\rho }}.\eqno{(3.20)}$$
It implies, in particular, 
$$
\sum_{i=1}^k U_i (\lambda ) U_{k+1-i} (\lambda)=
{1\over 2} U^\prime_{k+1} (\lambda ),\eqno{(3.21)}$$
where prime denotes differentiation with respect to the argument. 

For the normalization $N_s (\lambda )$ we need the 
following identities, also directly derivable from the 
recursion relations for the orthogonal polynomials:
$$
\sum_{i=1}^k U_i (\lambda ) U_i (\rho ) =
{1\over 2} {{U_{k+1} (\lambda ) U_k (\rho ) - 
U_{k+1} (\rho ) U_k (\lambda )}\over {\lambda -\rho }}.\eqno{(3.22)}$$
This implies, in particular,
$$
N^2 (\lambda ) ={1\over {\sum_{i=1}^k U_i^2 (\lambda ) }} =
{2\over{U^\prime_{k+1} (\lambda ) U_k (\lambda ) - U_{k+1} (\lambda ) 
U^\prime_k (\lambda ) }}.\eqno{(3.23)}$$

For the $M^{(2)}$ contribution we need: 
$$
\sum_{i=1}^k O_i (\lambda_1 (0)) O_{k+1-i} (\lambda_1 (0)) =
{1\over 2} N_1^2 (0) U^\prime_{k+1} (\lambda_1 (0)).\eqno{(3.24)}$$

For any $\lambda > 1$ and large $k$, $U_k (\lambda ) \sim (2\lambda)^{k-1}$
and $N(\lambda ) \sim {1\over {(2\lambda )^{k-1}}}$. Since
$\lambda_1 (0) >1$ we have proved the desired exponential suppression
for this contribution.

The first quantity in eq. (3.18) generates 8 
contributions from the two terms
in $D^{(1)}$ and the two terms in $G^{(0)}$. Only four of these
contributions commute with $\gamma_5$ and are of interest: 

The $W^{(1)}$--$W^{(1)}$ term will need the quantity defined below,
$$
f_{WW} (p) = \sum_{s=1}^k Y_s^2 (p) {{\mu_s (p)}\over
{\bar p^2 + \mu_s^2 (p)}},\eqno{(3.25)}$$
where 
$$
Y_s (p) = \sum_{j=1}^k O_j (\lambda_1 (0) ) 
O_j (\lambda_s (p)).\eqno{(3.26)}$$
The $W^{(1)}$--$M^{(1)}$ and $M^{(1)}$--$W^{(1)}$ terms are equal to
each other and for them we need the quantity
$$
f_{WM} (p) \equiv f_{MW} (p) = \sum_{s=1}^k X_s (p) Y_s (p) {1\over
{\bar p^2 +\mu_s^2 (p)}},\eqno{(3.27)}$$
where
$$
X_s (p) =\sum_{j=1}^k O_j (\lambda_1 (0) ) O_{k+1-j} (\lambda_s (p))
.\eqno{(3.28)}$$

For the $M^{(1)}$--$M^{(1)}$ we need
$$
f_{MM} (p) =\sum_{s=1}^k X_s^2 (p) {{\mu_s (p)}\over
{\bar p^2 + \mu_s^2 (p)}}.\eqno{(3.29)}$$

Each of the $f(p)$ functions goes into a loop integral 
over the four momentum $p$. Under the integral $f(p)$  is multiplied by
the explicit momentum dependence coming from the respective
vertices and by the gauge propagator which we simply 
take as 
$${1\over {2\sum_\mu (1-\cos p_\mu )}} \equiv {1\over {\hat p^2}};~~~~~~
\hat p_\mu = 2\sin {{p_\mu}\over 2}.\eqno{(3.30)}$$
Overall factors
coming from the Casimir in group space, $2\pi$'s in the integrals,
etc., are irrelevant to us here. 

It is now that we are in a position to see why something ``miraculous''
has to happen. If exponential suppression is maintained,
we expect every one of the contribution
to be suppressed individually, since, for
example, we can imagine varying $M^{(1)}$ without 
changing $W^{(1)}$, while maintaining the tree level mass hierarchy. 
Consider, for example, $f_{WW} (p)$. $Y^2_s (p)$ is bounded by 
1 via the Schwartz inequality. For arbitrary $p$ we would expect
the bound to be an order of magnitude estimate, and then we would
guess $f_{WW} (p)$ to be of order $k$. The smallness 
of $\mu_1 (p)$ for $p\in R$ hardly seems to make a difference.
To get a suppression the alternating
signs of the masses $\mu_s (p)$ must play a significant
role, inducing exponential suppression by almost perfect cancelations.
Similar considerations apply to the two other terms. 

Until now we have made no use what-so-ever of the 
structure of the vertices. We expect to need to 
use the fact that $M^{(1)}$ is linear in $p$ for small $p$,
since it is related to the RG irrelevance of the Wilson mass
term.

Let us start with the  $W^{(1)}$--$W^{(1)}$ term. From  
equations (3.4), (3.9), (3.22) we derive a simpler formula
for $N_s (p)$:
$$
{{N^2_s (p)}\over {\mu_s (p)}} ={2\over {b(p) 
U^\prime_{k+1} (\lambda_s (p) ) - U^\prime_k (\lambda_s (p) )}}.
\eqno{(3.31)}$$
Using now also equation (3.22)  we obtain 
$$
Y_s^2 (p) = {1\over {2 \mu_s (p)}} 
\left [ {{N_1 (0) U_k (\lambda_1 (0))}
\over {b (0)}} \right ]^2 \left [
{{b(0)-b(p)}\over {\lambda_1 (0) 
- \lambda_s (p)}} \right ]^2
{1\over{b(p) U_{k+1}^\prime (\lambda_s (p) ) 
- U_k^\prime (\lambda_s (p))}}
.\eqno{(3.32)}$$
It is now evident that for $s\ne 1$, $Y_s^2 (p)$ is indeed order 1, as 
estimated above. The sign carried by the $\mu_s (p)$ is also carried
by the denominator of the last term in (3.32). 
Inserting into eq. (3.25), and using (3.3), we get, for $p\ne 0$,
$$
\eqalign{
&f_{WW} (p) = {1\over 2}
\left [ {{N_1 (0)}\over {\mu_1 (0)}} \right ]^2 [b(p)-b(0)]^2\cr
&
\oint_{\cal C} {{dz}\over{2\pi i}} ~
{1\over {[\lambda_1 (0) - z ]^2}}~ {1\over {\bar p^2 + 1+b^2 (p) -2b(p) z}}
~{1\over{b(p)U_{k+1} (z)-U_k (z)}}.\cr}\eqno{(3.33)}$$
Here ${\cal C}$ encloses tightly all the zeros
of $\phi(z,k,p)\equiv b(p)U_{k+1} (z)-U_k (z)$. Along the real 
axis, $\phi(z,k,p)$ changes sign $k$ times and the cancelations
this causes are captured by deforming the contour of integration.
Deforming the contour to a circle
at infinity the line integral drops out and we pick up contributions
from the two extra poles $z_1 (k) = \lambda_1 (0)$ 
and $z_2 (p) 
={1\over 2} [{{{\bar p}^2}\over {b(p)}} +{1\over {b(p)}} + b(p) ]$. 
Both $z_1 (k)$ 
and $z_2 (p)$ are positive and larger than unity, so when plugged
into $\phi(z,k,p)$ give exponential suppression, 
as the prefactor, $ \left [ {{N_1 (0)}\over {\mu_1 (0)}} \right ]^2$,
is of order one. 
Moreover, $0<z_1 (k) -1$ is bounded away from zero
for all $k$, while $0<z_2 (p) -1$ is bounded away from zero 
for all $p$. 

However, the exponential suppression is wiped out at very 
small $\hat p^2 $ since, 
by the definition of $\lambda_1 (0)$, $\phi (z_1 (0),k,0) =0$. The region of
very small $p$ where we have no exponential suppression is
best analyzed using the original expressions (3.25) and (3.26).
Since $Y_s (0) =\delta_{1s}$ by ortho-normality, the dominating term
at very small momenta is ${{\mu_1 (p)}\over {\bar p^2 +\mu_1^2 (p)}}$.
When multiplied by the propagator (3.30) we see that the suppression
factor $ \mu_1 (0)$ will be multiplied by $ \log \mu_1^2 (0)$. This
behavior is expected from the continuum, and indeed the small $p$
region of the loop integral is the place where continuum perturbation
theory is reproduced ([\smitold]). We learn that we should expect 
extra multiplicative factors of order $k$ in front of the exponentially
suppressed term. Here, we basically ignore these factors.
The region of small $\hat p^2$ can be taken to extend
up to $\hat p^2 \sim |\mu_1 (0)|^a$  with $0<a<1$
and the contribution from the rest of the integral is then seen to be
exponentially suppressed. 

We now turn to the $W^{(1)}$--$M^{(1)}$ and $M^{(1)}$--$W^{(1)}$ terms.
Manipulations similar to the above produce:
$$
X_s (p) Y_s (p) = {{N^2_1 (0)}\over {2\mu_1 (0)}}~
{{b(p)-b(0)}\over {[\lambda_1 (0) -\lambda_s (p) ]^2}}~
{{U_{k+1} (\lambda_1 (0)) - U_{k+1} (\lambda_s (p)) }\over
{b(p) U_{k+1}^\prime (\lambda_s (p) ) 
- U_k^\prime (\lambda_s (p))}}.\eqno{(3.34)}$$
Again, we represent $f_{WM} (p)$ by
$$
\eqalign{
&f_{WM} (p) = {{N^2_1 (0)}\over {2 \mu_1 (0)}}  [b(p)-b(0)] \cr
&\oint_{\cal C} {{dz}\over{2\pi i}} ~
{{U_{k+1} (\lambda_1 (0) ) - U_{k+1} (z)}\over 
{[\lambda_1 (0) - z ]^2}}~ {1\over {\bar p^2 + 1+b^2 (p) -2b(p) z}}
~{1\over{b(p)U_{k+1} (z)-U_k (z)}}.\cr}\eqno{(3.35)}$$

The contour is first defined as above, and then deformed as before to infinity.
Again the line integral does not contribute, and we end up only with
contributions from the poles at $z_1 (k)$ and $z_2 (p)$. 
From $z=z_1 (k)$ and momenta not too small
we obtain a contribution of order one from the
integral and the 
prefactor ${{N^2_1 (0)}\over {\mu_1 (0)}}$ provides exponential suppression.
From $z=z_2 (p)$ we obtain a contribution which, as a result 
of $U_{k+1} (\lambda_1 (0)) << U_{k+1} (z_2 (p))$ for not
too small $p$ and $k$ large enough, is also of order one. Thus the
overall exponential suppression is, using (3.31), by
$$
{{N^2_1 (0)}\over {\mu_1 (0)}}\sim  [ 2z_1 (k) ]^{-k}.\eqno{(3.36)}$$

For $p\to 0$ (3.27) will be dominated by the $s=1$ term, given 
by $X_1 (0) {1\over {\bar p^2 + \mu^2_1 (p) }}$. In addition we have a 
factor of ${1\over {\hat p^2}}$ from the gauge field propagator and
a factor going as $\hat p^2$ from the $M^{(1)}$ vertex combined with
the $\gamma_\mu \bar p_\mu$ term from $G^{(0)} (p)$. Thus, the region
of momenta where (3.35) is not suppressed is too small to eliminate
the exponential suppression found for momenta away from zero. 
Note that $X_1 (0)$ is also exponentially small. 

The last term to be analyzed is of type $M^{(1)}$--$M^{(1)}$.
We need the following expression, derived with the help of (3.20),
$$\eqalign{
X^2_s (p) =& {1\over 4} N^2_s (p) N^2_1 (0) \left [
{{U_{k+1} (\lambda_1 (0)) - U_{k+1} (\lambda_s (p)) } \over
{\lambda_1 (0) - \lambda_s (p) }} \right ]^2\cr
X^2_1 (0) =& {1\over 4} \left [ N_1^2 (0) 
U^\prime_{k+1} (\lambda_1 (0))\right ]^2 .\cr}\eqno{(3.37)}$$
With it we obtain:
$$
\eqalign{
f_{MM} (p) = {1\over 2} N^2_1 (0) \sum_{s=1}^k & {{\mu^2_s (p)}\over
{\bar p^2 + 1 + b^2 (p) -2b(p) \lambda_s (p)}}~
{1\over{b(p)U^\prime_{k+1} (\lambda_s (p)) - U^\prime_k (\lambda_s (p))}}\cr
&\left [ {{U_{k+1} (\lambda_1 (0)) - U_{k+1} (\lambda_s (p))}\over
{\lambda_1 (0) - \lambda_s (p)}} \right ]^2 .\cr}\eqno{(3.38)}
$$
From (3.29) we see that $f_{MM} (p)$ has a finite limit of order one
as $\hat p^2\to 0$. In the Feynman diagram the gauge propagator singularity
at $\hat p^2\to 0$ is  canceled by the momentum dependence coming from
the $M^{(1)}$ vertices. Therefore, we can assume 
that $p_\mu$ is sufficiently far from zero to permit 
individual treatment of the terms obtained from an expansion of 
the binomial squared in the numerator of the last factor in (3.38).
There are three terms. The most ``dangerous'' contains the $s$-independent
large constant $U^2_{k+1} (\lambda_1 (0))$ (see below
for the other two terms). The contribution from
this term is:
$$\eqalign{{1\over 2}
N^2_1 (0) U^2_{k+1} (\lambda_1 (0))& \oint_{\cal C} {{dz}\over{2\pi i}}\cr ~
&{1\over {[\lambda_1 (0) - z ]^2}}~ {
{1+b^2(p) -2 b(p) z}\over {\bar p^2 + 1+b^2 (p) -2b(p) z}}
~{1\over{b(p)U_{k+1} (z)-U_k (z)}}.\cr}\eqno{(3.39)}$$
The prefactor is of order unity but the integral is 
exponentially suppressed as in the cases above. 
The other two terms, are also exponentially suppressed:
The cross term has an exponentially small prefactor
and an integral of order one. In the last term
one needs to replace in the sum over $s$ the 
factor $U_{k+1}^2 (\lambda_s (p))$ 
by ${1 \over {1+b^2 (p) -2b(p) \lambda_s (p)}}$
using equation (3.11). This avoids too fast growth
at complex infinity in the associated contour integral.
Now, one sees the exponential suppression easily.

This concludes the argument. Note that a gauge propagator which
is more singular in the infrared would have invalidated our
argument by enhancing the relevance of the very low momentum
region in the diagram. It would be hard to see this directly
from the analysis of section 2, and this provides one
example where the calculations of section 3 are seen to be
explicitly needed. 
If one needs explicit, complete formulae
the above analysis applies because all sums over $s$ can be
evaluated in closed form with the help of the complex contour integrals
shown. A similar technique was used in [\kikkaw]. All we do applies
directly also to the ``almost'' supersymmetric case discussed in
[\recentb] by methods similar to [\huetnn].
\vskip 1cm
\centerline{\bf 4. Free fermion propagator.}
\vskip .5cm
Using the methods of this paper we derive first a formula for the free
propagator. We shall make the formula completely explicit for
the Chebyshev case, but we wish to present it in a less explicit
manner first because, at that stage, 
the structure would be the same for other sets of orthogonal
polynomials.

According to (3.14) (see also [\plbfirst]) we only need to invert
the second order Dirac operator. So we need the matrix
$$
\Delta ={1\over {\bar p^2 + m^2 (p)}}=O(p) {1\over{\bar p^2 +\mu^2 (p)}}
O^T (p).\eqno{(4.1)}$$
Explicitly,
$$
\Delta_{ij} (p) = \sum_{s=1}^k N^2_s (p) 
{{U_i (\lambda_s (p)) U_j (\lambda_s (p))}\over {\bar p^2 +\mu_s^2 (p)}}.
\eqno{(4.2)}$$
Although not indicated explicitly, $N_s$ also depends on $k$.
Using the same techniques as before we arrive at
$$
\Delta_{ij} (p) =2b(p) \oint_{\cal C} {{dz}\over {2\pi i}}
{{U_i (z) U_j (z)}\over {U_k (z) [b(p) U_{k+1} (z) - U_k (z)]}}
{1\over {\bar p^2 +1 +b^2 (p) -2b(p) z}}.\eqno{(4.3)}$$
Here the contour ${\cal C}$ encircles precisely only the
roots of $b(p) U_{k+1} (z) - U_k (z)$. Deforming to infinity
we pick up one contribution from the pole at $z=z_2 (p)$ 
and $k-1$ contributions from the roots of $U_k (z)$, denoted by $x_t$:
$$\eqalign{
\Delta_{ij} (p) =& -2 \sum_{t=1}^{k-1} 
{{U_i (x_t ) U_j (x_t )}\over {U^\prime_k (x_t ) 
U_{k+1} (x_t )}}
{1\over {\bar p^2 +1 +b^2 (p) -2b(p) x_t }}+\cr
& {{U_i (z_2 (p) ) U_j (z_2 (p) )}
\over {U_k (z_2 (p) ) [b(p) U_{k+1} (z_2 (p) ) - U_k (z_2 (p) )]}}.\cr}
\eqno{(4.4)}$$
The first term will be written as  $\Delta^0_{ij} (p)$ while the
second is, up to normalization, a projection matrix on a state
that, for $p\in R$, 
has entries exponentially decreasing with the distance of $i$
from $k$. At $k$, the components are order unity. The second term
represents the contribution of a state dominated by
flavors near $k$. This state clearly is representing the almost
massless quark. On the other hand  $\Delta^0_{ij} (p)$ vanishes
if either $i$ or $j$ is equal to $k$. Thus, it represents some other
set of $k-1$ states. Jointly, these states span a $k-1$ subspace in
flavor space. This {\it subspace} is not dependent on the momentum $p$.

Using the recursion relations we see that we can replace 
in the denominator the quantity $U_{k+1} (x_t )$ 
by $-U_{k-1} (x_t )$. Introducing now the orthogonal 
matrix $O^{(k-1)} $ ($p$ independent) defined as
$$
\eqalign{
O^{(k-1)}_{it}  =& N_t^{(k-1)} U_i (x_t)\cr
(N_t^{(k-1)})^2 =&{1\over {\sum_{i=1}^{k-1} U_i^2 (x_t)}},\cr}\eqno{(4.5)}$$
we see that, for $i,j=1,...k-1$,
$$\eqalign{
\Delta^0_{ij} (p)=&\sum_{t=1}^{k-1} 
{{ O^{(k-1)}_{it} O^{(k-1)}_{jt}}\over
{\bar p^2 +\mu^2_t (p)}}=
\left ( {1\over {\bar p^2 + m_0^2 (p)}} \right )_{ij},\cr
\mu_t^2 (p)  =& 1+ b^2 (p) -2b(p) x_t ,\cr
m_0^2 (p) =& 1 +b^2 (p) -2 b(p) {\bf J} ,~~~~~~~
{\bf J} = \pmatrix { 0 & 1 & 0 & 0& \ldots &0&0\cr
		     1 & 0 & 1 & 0 &\ldots &0&0\cr
		     0 & 1 & 0 & 1 &\ldots &0&0\cr
		     \vdots &\vdots &\vdots &\vdots&\vdots & \vdots&\vdots\cr
      \ldots & \ldots & \ldots & \ldots &\ldots & 0 & 1\cr
		     \ldots & \ldots & \ldots & \ldots &\ldots & 1 & 0\cr } 
.\cr}\eqno{(4.6)}$$

Above, the matrix ${\bf J}$ has dimensions $(k-1)\times (k-1)$. 
The form of the ${\bf J}$ term in the
matrix $m_0^2 (p)$ was determined by considering what matrix
would have the eigenvalues $x_t,~t=1,2,...,k-1$, by analogy with
[\recentb] where the inverse problem was solved. 
Note that although we defined the matrix $m_0^2 (p)$ it does not have
a simple (sparse) square root, unlike $m^2 (p)$. All $k-1$ eigenvalues
of $m_0^2 (p)$ are positive and bounded away from zero for
all momenta $p$ because all the zeros of $U_k (z)$ are
in the interval $(-1,1)$ for any $k$.

It is clear that the above separation of the second order operator
differentiates between the contributions of the light Wilson 
fermion and the other $k-1$ heavy Wilson fermions. In
flavor space the heavy Wilson fermions span a fixed $k-1$ dimensional
subspace. The light Wilson fermion is associated with a direction
that is almost orthogonal to the heavy sector of flavor space
for $p\in R$. The heavy
doublers contained in the light Wilson fermion appear to have 
unsuppressed overlaps with the heavy Wilson fermions, but some
cancelations prevent them from giving substantial mass to the
light component of the light Wilson fermion. 

The full propagator is easily obtainable now, once the explicit
form of $m (p)$ is used. 

To this point we only did manipulations that we believe would generalize
to other sets of orthonormal polynomials. We now write down the
answers allowing ourselves to use specific properties of the Chebyshev
polynomials. In [\taniguchib] similar quantities where calculated following
steps laid out in [\shamir] who 
generalized the method of [\plbfirst]
to the finite $k$ case. 

Following [\plbfirst] we introduce the positive 
quantity $\alpha (p)$:
$$
z_2 (p)
={1\over 2} \left [ 
{{{\bar p}^2}\over {b(p)}} +{1\over {b(p)}} + b(p) \right ] 
\equiv \cosh (\alpha(p)),~~~b(p) = M_0 + {1\over 2} \hat p^2,~~~0<M_0 <1. 
\eqno{(4.7)}$$
It is important to note ([\plbfirst])
that $\alpha(p)$ is smooth and bounded away from zero
for all momenta $p$. 
Using standard manipulations of trigonometric
identities and contour integrations, we derive in the 
appendix the following explicit
expression for the free propagator, including all finite $k$ effects:
$$\eqalign{
\Delta_{ij} (p) = &
{{\sinh \{ \alpha (p) [ \min (k-i,k-j) ] \} \sinh \{ \alpha (p) 
[\min (i,j)]
\}}\over{b(p) \sinh [\alpha (p) ] \sinh [k\alpha (p)]}} +\cr
&{{\sinh[i\alpha (p)]\sinh [j\alpha (p) ]}\over 
{\sinh [k\alpha (p) ] \{ b(p)\sinh [(k+1)\alpha (p) ] -
\sinh [k\alpha (p)] \}}}.\cr}\eqno{(4.8)}$$
The term on the second line in (4.8) contains the light particle. 
One large $k$ limit is obtained by taking $k$ to infinity with
$k-i\equiv i^\prime $ and $k-j\equiv j^\prime $ held fixed. 
In this limit, $i^\prime, j^\prime =0,1,2....\infty$, and we obtain,
for $\hat p^2 \ne 0$,
$$
\Delta^{(1)\infty}_{ij} (p) ={{e^{-\alpha(p) |i^\prime -j^\prime | } -
e^{-\alpha(p) (i^\prime + j^\prime )} }\over
{b(p)\left [ e^{\alpha(p)} -e^{-\alpha(p)} \right ]}} 
+ {e^{-\alpha(p) (i^\prime + j^\prime )}\over
{ b(p) e^{\alpha(p)} - b(0) e^{\alpha (0)} }}
.\eqno{(4.9)}$$
Note that $b(0)\exp [\alpha(0)]\equiv 1$, but when 
written as above, the pole at $\hat p^2 =0$ becomes evident. 
As $k\to\infty$, for $\hat p^2 =0$ and fixed, finite $i^\prime, j^\prime$, 
the second term diverges as $\left ( {1\over {b(0)}} \right )^{2k} $,
reflecting the exponential smallness of the mass. 
Another large $k$ limit is obtained 
taking $k\to\infty$ with $i$ and $j$
kept finite, $i,j=1,2,....$. The second term in (4.8) 
disappears for $\hat p^2 \ne 0$ while the first is similar 
to (4.9) because of its invariance under $i\to k-i,~j\to k-j$:
$$
\Delta^{(2)\infty}_{ij} (p) = 
{{e^{-\alpha(p) |i -j | } -
e^{-\alpha(p) (i + j )} }\over
{b(p)\left [ e^{\alpha(p)} -e^{-\alpha(p)} \right ] }} .\eqno{(4.10)}$$
Again, for $\hat p^2 =0$, special analysis is required.  
In the two above 
limits we choose to keep two distinct groups of particles.
These two groups decouple at infinite $k$.
Exponential mass suppression requires this decoupling to hold
approximatively also at finite $k$. 
In both cases the 
limits $\hat p^2 \to 0$ and $k\to\infty$ do not commute. The regime
where this lack of commutativity is felt can eventually become
important if the gauge propagator is replaced by something with
a stronger divergence in the infrared. For the ordinary singularity,
in four dimensions, 
the lack of commutativity appears to have no major effect to one loop 
order. However, in two dimensions, the ordinary singularity is
already sufficiently strong to require a deeper investigation.

The full Dirac propagator is  given by:
$$
G^{(0)} (p) = \Delta (p) [ -i\gamma_\mu \bar p_\mu + m(p) ] \equiv
[ -i\gamma_\mu \bar p_\mu + m(p) ] \Delta (p),\eqno{(4.11)}$$
where the $k\times k$ matrix $m(p)$ has the following structure:
$$
m=\pmatrix{0&0&\ldots&0&-1&b(p)\cr
	   0&0&\ldots&-1&b(p)&0\cr
	   \vdots&\vdots&\vdots&\vdots&\vdots&\vdots\cr
	   \vdots&\vdots&\vdots&\vdots&\vdots&\vdots\cr
	   -1&b(p)&\ldots&\ldots&0&0\cr
	   b(p)&0&\ldots&\ldots&0&0\cr}.\eqno{(4.12)}$$

If one wishes to keep a finite mass at $k=\infty$ it has to
be introduced by hand [\npblong ,\shamir ]. Since it is obvious
that the flavor $k$ strongly mixes with the light fermion,
one can use this component of the fermions as an interpolating
field for the light particle [\plbfirst]. Even at finite $k$
one may find some advantages to keep such an explicit extra
mass term [\recentb]. In our notation it amounts to 
replacing the mass matrix $m(p)$ by $m(p) +\mu {\bf N}$, 
where $\mu$ is the new mass parameter and ${\bf N}$ has
a single nonzero entry: ${\bf N}_{ij} =\delta_{ik} \delta_{jk}$,
so as to couple only the left and right components of
the interpolating field. 

In equation (4.9) we have the propagator for $\mu =0$. It
is easy to obtain the propagator for  $\mu\ne 0$, $G^{(0)}_\mu (p)$:
$$
G^{(0)}_\mu (p) = {1\over{i\gamma_\mu \bar p_\mu + m(p) + \mu {\bf N}}}
.\eqno{(4.13)}$$
Since ${\bf N}$ obeys
$$
{\bf N}^2 = {\bf N},~~~~~~~{\bf N}G^{(0)} (p) {\bf N} = g_0 (p) {\bf N},
\eqno{(4.14)}$$
where $g_0 (p) = G^{(0)}_{kk} (p)$, one can simply expand in $\mu$
and resum the series:
$$ \eqalign{
G^{(0)}_\mu (p) &= G^{(0)} (p) -\mu  G^{(0)} (p) {\bf N} G^{(0)} (p)
+\mu^2  G^{(0)} (p) g_0 (p) {\bf N} G^{(0)} (p) -\cr\mu^3 
 &G^{(0)} (p) g_0^2 (p) {\bf N} G^{(0)} (p)+......= 
 G^{(0)} (p) -  G^{(0)} (p) {{\mu}\over {1+\mu g_0 (p) }} 
 {\bf N} G^{(0)} (p).\cr}
\eqno{(4.15)}$$
Note that as far as spinorial indices go, $g_0 (p)$ is still a matrix,
and ordering is important. 
So, the free propagator with the extra mass term $\mu$ is given by:
$$
{G^{(0)}_\mu (p)}_{ij} 
= G^{(0)}_{ij} (p)  - G^{(0)}_{ik}(p) 
{{\mu}\over {1+\mu G^{(0)}_{kk}(p) }}
G^{(0)}_{kj}(p).\eqno{(4.16)}$$
Since $\psi_k$ is the interpolating field for the light fermion
one can imagine integrating out all the other fermion fields 
[\plbfirst]. The action, for $\mu=0$, would obviously 
be $\bar\psi_k {1\over {G_{kk}^{(0)}}} \psi_k$. Now the introduction
of $\mu\ne 0$ has a trivial effect, merely adding the 
term $\mu\bar\psi_k \psi_k$. Therefore, the new free propagator for 
the $\psi_k$ field is:
$$
{G^{(0)}_\mu (p)}_{kk} = {1\over{ {1\over {G_{kk}^{(0)}(p)}}+\mu}}
.\eqno{(4.17)}$$
It is easy to see that this coincides with eq. (4.16). 
$G_{kk}^{(0)}(p)$ has a relatively simple form:
$$
G_{kk}^{(0)}(p)={{-i\gamma_\mu \bar p_\mu \sinh [k\alpha (p) ]
+b(p) \sinh [\alpha (p)] }\over
{b(p)\sinh [(k+1)\alpha (p) ] -\sinh [k\alpha (p) ]}}.\eqno{(4.18)}$$

It is unclear at the present whether employment of the free propagator
given above would result in a simplified analysis in the truncated
overlap case. 
Our hope is that the more general treatment we presented
in section 3
will promote a search for more efficient mass suppression schemes. Note
that the analysis in section 2 relied on the particular structure of the
truncated overlap from an earlier stage. Nevertheless, it too can
be made more general, using the formulae given in the appendix of
[\recentb]. 
\vfill\eject
\centerline{\bf 5. Summary and Outlook.}
\vskip .5cm

The evidence in favor of exponential suppression of the light quark
mass in the truncated overlap and potentially in related systems is fairly
strong. Thus there is a new competitor to existing methods
designed to reduce/eliminate ${\cal O} (a)$ effects in QCD simulations.
In practice, the requirement $M_0 > 0$ cannot be met at practical
values of the gauge coupling. It remains to be understood whether
the required negative values of $M_0$, and the potentially associated
``exceptional configurations'' are better, worse or similar to those
affecting ordinary Wilson fermions. 
In [\recentb] it was suggested that for $-1< M_0 < 0$  and odd $k$'s
one would be dealing with the $\theta = \pi$ regime of $QCD$. It would be
interesting to see how much of the present paper extends to this case.
Whether this suggestion should extend to the $k=1$ case (ordinary
Wilson fermions) or not is too early to even guess.

\vskip .1in

\noindent {\bf Acknowledgments}:
Y.K. would like to thank T.~Kugo for enlightening discussions.
The research of HN was supported in part by the 
DOE under grant  \# DE-FG05-96ER40559. HN wishes to
thank Pavlos Vranas for an informative exchange. 
AY would like to thank Amol Dighe, Marco Moriconi, George Thompson 
and Andrija Rasin for discussions. 

\vskip 1cm
\centerline{\bf 6. Appendix.}
\vskip .5cm

This appendix deals exclusively with the Chebyshev case. The
results therefore are not expected to generalize easily
to other sets of orthogonal polynomials. Of course, the
truncated overlap (domain wall) is covered.

For Chebyshev polynomials we have:
$$
U_k (x) = {{\sin (k \theta )}\over \sin (\theta)};~~~~~x=\cos(\theta )
.\eqno{(A.1)}$$
Hence,
$$
x_t \equiv \cos (\theta_t );~~~~\theta_t = t{\pi\over k};~~~~t=1,....,k-1.
\eqno{(A.2)}$$
A short calculation then gives:
$$\eqalign{
\left ( N_t^{(k-1)} \right )^2 =& 
{2\over {U^\prime_k (x_t ) U_{k-1} (x_t )}}
={2\over k} \sin^2 (\theta_t ),\cr
O^{(k-1)}_{it} O^{(k-1)}_{jt} =& {1\over k} \left [
\cos ((i-j)\theta_t ) - \cos ((i+j)\theta_t ) \right ].\cr}\eqno{(A.3)}$$
Thus,
$$
\Delta^0_{ij} (p) ={1\over {kb(p)}} \sum_{t=1}^{k-1} 
{{\cos \left ( {{i+j}\over k} t\pi \right ) - 
\cos \left ( {{i-j}\over k} t\pi \right )}\over
{2\cos \left ( {{t\pi}\over k}\right ) - 
2\cosh (\alpha (p))}}.\eqno{(A.4)}
$$
Replace $t$ by $k-t$ in the above sum and average the two expressions.

For $i+j$ even we obtain 
$$
\Delta^0_{ij} (p) = 2\cosh (\alpha (p)) {{1}\over {kb(p)}}\sum_{t=1}^{k} 
{{\cos \left ( {{i+j}\over k} t\pi \right ) - 
\cos \left ( {{i-j}\over k} t\pi \right )}\over
{\left [ 2\cos \left ( {{t\pi}\over k}\right ) \right ]^2
- \left [ 2\cosh (\alpha (p)) \right ]^2 }}.\eqno{(A.5)}
$$
Note that we have extended the sum to $t=k$ because the 
corresponding term in the sum vanishes. Introduce now two
non-negative integers, $l$, $l^\prime$:
$$
l={{i+j}\over 2},~~~~~~~ l^\prime = {{|i-j|}\over 2}.\eqno{(A.6)}$$
$\Delta^0_{ij} (p)$ is seen to be expressible in terms 
of a quantity $g_l (p)$,
$$
g_l (p) \equiv \sum_{t=1}^k {{\cos \left (2\pi t {l\over k} \right )}
\over {\left [ 2\cos \left ( {{t\pi}\over k}\right ) \right ]^2
- \left [ 2\cosh (\alpha (p)) \right ]^2 }} =
{\rm Real } \left ( \sum_{t=1}^k {{z_t^{1-l}}\over { \left [
z_t - e^{2\alpha (p)} \right ] \left [
z_t - e^{-2\alpha (p)} \right ] }}\right ), \eqno{(A.7)}$$
where $z_t = e^{2i\pi {t\over k}}, ~t=1,...,k$ are all the solutions
of $z^k=1$. 

Consider now a complex line integral $I$ over a closed circle 
at infinity:
$$
I=\oint {{dz}\over {2\pi i}} {{kz^{k-l}}\over {z^k -1}}
\left [ {1\over {z - e^{2\alpha (p)}}} - {1\over {z - e^{-2\alpha (p)}}}
\right ].\eqno{(A.8)}$$
$I$ vanishes for any $l\ge 0$. This leads to
$$
g_l (p) = -{k\over 2}{{ \cosh [ (k-2l ) \alpha (p) ] }\over
{\sinh [2\alpha (p) ] \sinh [k \alpha (p) ]}},\eqno{(A.9)}$$
from which $\Delta^0_{ij} (p)$  can be obtained for even $i+j$.

For odd $i+j$ we define 
$$
l={{i+j+1}\over 2},~~~~~~l^\prime ={{|i-j|+1}\over 2}.\eqno{(A.10)}$$
Note that now $l,l^\prime \ge 1$. A few lines lead us to
$$
\Delta^0_{ij} (p) = {1\over {kb(p)}} 
\left [ g_l (p) + g_{l-1} (p) - g_{l^\prime } (p) - g_{l^\prime -1} (p)
\right ].\eqno{(A.11)}$$

Collecting all results we see that the expressions for even $i+j$ 
and odd $i+j$ are the same and our final result becomes:
$$
\Delta^0_{ij} (p) =  
{{\cosh \left [ (k-|i-j| )\alpha (p) \right ]
- \cosh \left [ (k-i-j) \alpha (p) \right ]}\over
{2b(p) \sinh [\alpha (p) ] \sinh [k\alpha (p) ]}}.\eqno{(A.12)}$$
A few more steps produce equation (4.8).

\vfill\eject

\noindent{\bf References}
\vskip .1in
\item {[\npblong]} R. Narayanan, H. Neuberger, Nucl. Phys. B443 (1995) 305.
\item {[\recent]} H. Neuberger, hep-lat/9707022, Phys. Lett. B, to appear.
\item {[\recentb]} H. Neuberger, hep-lat/9710089.
\item {[\blumsoni]} T. Blum, A. Soni, Phys. Rev. D56 (1997) 174. 
\item {[\vranas]} P. Vranas, hep-lat/9705023, hep-lat/9709119.
\item {[\plbfirst]} R. Narayanan, H. Neuberger, Phys. Lett. B302 (1993) 62.
\item {[\kaplan]} D. B. Kaplan, Phys. Lett. B288 (1992) 342.
\item {[\shamir]} Y. Shamir, Nucl. Phys. B406 (1993) 90.
\item {[\seesaw]} M. Gell-Mann, P. Ramond, R. Slansky 
in {\it Supergravity} edited by P. van Nieuwenhuizen 
and D. Z. Friedman (North-Holland, 1979).
\item {[\fn]} C. D. Froggatt,  H. B. Nielsen, Nucl. Phys. B147 (1979) 277.
\item {[\taniguchi]} S. Aoki, Y. Taniguchi, hep-lat/9709123.
\item {[\taniguchib]} S. Aoki, Y. Taniguchi, hep-lat/9711004.
\item {[\yamada]} A. Yamada, hep-lat/9705040, Phys. Rev. D, to appear; 
hep-lat/9707032, Nucl. Phys. B, to appear.
\item {[\npold]} R. Narayanan, H. Neuberger, Nucl. Phys. B. 412 (1994) 574.
\item {[\smitold]} L. H. Karsten, J. Smit, Nucl. Phys. B188 (1981) 103.
\item {[\kikkaw]} T. Kawano, Y. Kikukawa, Nucl. Phys. 
B(Proc. Suppl.)47 (1996) 599, \hfill\break hep-lat/9501032. 
\item {[\huetnn]} P. Huet, R. Narayanan, H. Neuberger, Phys. 
Lett. B380 (1996) 291.

\vfill\eject
\end